# FLBRA: Fuzzy Logic Based Routing Algorithm for Indoor Wireless Sensor Networks

Lucas Leão[1], David Bianchini and Omar Branquinho

Pontifical Catholic University of Campinas, Campinas, Brazil

## ABSTRACT

*Considering the context of building management systems with wireless sensor networks monitoring environmental features, this paper presents a proposal of a Fuzzy Logic Based Routing Algorithm (FLBRA) to determine the cost of each link and the identification of the best routes for packet forwarding. We describe the parameters (Received Signal Strength Indicator - RSSI, Standard Deviation of the RSSI and Packet Error Rate - PER) for the cost definition of each path, the sequence of identifying best routes and the results obtained in simulation. As expected in this proposal, the simulation results showed an increase in the packet delivery rate compared to RSSI-based forward protocol (RBF).*

## 1. INTRODUCTION

Wireless sensor networks (WSN) have been a widely used solution for monitoring systems, including, for example, Building Management Systems (BMS). The sensors are responsible for monitoring aspects of the environment, capturing data such as temperature, lighting and power consumption [1]. Through this monitoring, management strategies and control of lighting, and HVAC (heating, ventilation, and air conditioning) are applied seeking greater efficiency in resource use.

However, when it comes to an application of WSN in a dynamic environment, such as offices, subject to constant physical changes, either due to the inclusion of new obstacles like furniture and walls as well as the movement of people throughout the space, the design of the wireless network should take into consideration the possibility of constant reconfiguration of the network. These reconfigurations are necessary since such changes in the environment generate impacts on the initial performance of the WSN [2].

In this scenario, it is important to identify solutions to meet the demands of reliability and to overcome the indoor environment constraints. To do so, in order to minimize the impacts of changes in the environment, this paper presents a strategy for network reconfiguration, by dynamically changing the routes of communication between the sensors, aiming the continuous communication between the sensors and the base station. The proposal makes use of fuzzy logic to assign costs to links considering the relation between RSSI, Standard Deviation of the RSSI, and Packet Error Rate.

The organization of this paper includes the description of related work, describing routing solutions based on RSSI in Section 2. Section 3 describes a brief background on Fuzzy Logic. The description of our solution, with the proposed algorithm, is shown in Section 4. Section 5 presents the discussion of the simulation results. Finally the conclusion in Section 6.

## 2. RELATED WORKS

Routing solutions based on RSSI are characterized by the use of RSSI measured values between sensor nodes and the base station to define the routes to be used for packet forwarding [3]. Such protocols may consider the physical distance between the sensor node and the base station, choosing the closest sensor and electing it as responsible for forwarding packets to the base station from more distant sensors [4]. Despite of that, it is risky to assume that sensor nodes with high level of RSSI are near the base station, since shadowing effect may generate unpredicted outcomes [3].

The proposal presented in [3,4] is a data delivery protocol that combines decisions in MAC and routing layers in order to determine the next hop in the forwarding process. It is an event-driven protocol, with distributed processing and state-free design. There is no need for the sender to know the network topology or its neighbour's location. Differently from the location-based protocols, RBF is unaware of sink location, and the forwarding is based on the measured RSSI received from the sink. It is assumed that the sink is capable of sending beacons, which can be heard by all sensor nodes in the network. Based on that, sensors are expected to use the RSSI as a parameter to determine the closest sensor to the node. The strategy goes as a competition based selection, where the sensor node with the highest RSSI level is selected to forward the data to next hop. Since packet delivery rate is not a concern in this strategy, it is not assured that the next hop is the best choice and it will effectively deliver the forwarded packet (void areas).

Reference [5] presents a decentralized packet forwarding solution based on RSSI. It is a location-free greedy forward algorithm, which makes use of broadcast messages to decide the next hop of the forwarding task. The authors describe two different approaches, namely RSSR Election and RSSR Selection. The first approach is based on the election of a leader, which will hold the task of forwarding the packet. In this solution, the sink node, which is assumed to have a powerful radio range capable of reaching the entire network, sends a query to a specific node in the form of a broadcast message. Then, all nodes in the network saves the RSSI from the sink in a local database. After checking if the query is addressed to itself, the node sends a broadcast message to all its neighbours. The neighbour having the closest distance from the sink (calculated from the RSSI) elects itself as the leader and forwards the packet in a form of a broadcast message. The other nodes participating in the election receive the message and cancel their participation. The second approach runs differently by not starting an election, but triggering a selection phase. After receiving the message from the sink, nodes start exchanging advertisements, so nodes can be aware of the distance of their neighbours from the sink. The information is stored in a local routing table, which is then used by the inquired node to select the next hop (the neighbour closest to the sink). Notwithstanding the gains of a distributed solution, such as implicit reaction to network changes, the proposed solutions do not address issues related to voids in the network, resulting in packet not being delivered.

In [6] a solution based on the characteristics of signal propagation is presented. The authors' proposed algorithm is a centralized solution that checks the status of the links between sensor nodes and builds a routing table, which is stored within each sensor node. Then it starts to monitor the links, noting possible degradation in the signal. If there is a link in the routing table whose parameters are below a desired threshold, the algorithm reconstructs the routing table, identifying the links again with values within the required limit. The proposal uses the packet delivery rate and signal to noise ratio (SNR) of each link as a parameter to evaluate the eligibility of the route as an entry in the route table. This solution is an improvement of a previous work [7] presenting a static routing protocol that used the average of measured RSSI and the packet delivery ratio as parameters to construct the routing table. In both proposals, authors define a relation between link quality and packet delivery ratio in order to classify the links of each sensor and create a routing table. Although it is a dynamic solution, capable of adapting itself to changes in the link quality, it is still subject of voids in the network, since it establishes thresholds for the measured parameters. It means that occasionally sensor nodes may be out of the defined threshold and consequently out of the routing table, creating voids in the network.

The method suggested by [8] makes use of Fuzzy Logic to select the sensor nodes that will be used for the packet forwarding task. It considers two main variables in order to select the next hop: sensor node angle in relation to its neighbours and number of packets forwarding to the neighbour. Those variables are analysed through a fuzzy system and the outcome is a number that represents the chance of a sensor node being selected as next hop. The next step consists of selecting the neighbour node which have the highest chance to forward the packet. Although it is a distributed solution, sink location must be known by sensor nodes in order to correctly select the next hop. Location is obtained by comparing the angle of the node in relation to the sink. It implies that the base station cannot have its location changed; otherwise the entire network would be lost. Moreover, there is no controlling method to avoid nodes with high rates of packet loss.

In [9] the authors propose a decentralized algorithm based on TARF (Trust-Aware Routing Framework) to determine the reliability of a sensor within the network and to verify if the information from the sensor nodes are eligible for aggregation or not, thus defining the routes for packet forwarding. The proposal makes use of Fuzzy Logic as a tool to support the decision process. Variables such as Energy Cost, distance, signal strength and packet delivery ratio, which after a process of fuzzification, serve as input to the decision rules.

Another proposal is presented in [10] describing a method for route selection using fuzzy logic and considering the evaluation of three parameters: battery level, trust level and distance from the base station considering the RSSI level. The value of the residual battery level is calculated taking into account the energy used for transmission at a given distance. The proposed model takes into account a decentralized solution to select nodes that should serve as a path in the routing process. It is still considered the existence of a database, in the sensor node itself, for storing information about the history of neighbour sensor nodes.

In [11] it is presented an estimator of the link quality between sensor nodes using Fuzzy Logic. Parameters such as packet delivery rate, link asymmetry in uplink and downlink, the stability of the link (changes in the measured parameters) and channel quality (signal to noise ratio) are used as metrics for the link quality evaluation. The authors justify the use of Fuzzy Logic in the link quality estimation due to the imprecision in the measures of the factors that affects the quality and stability of the communication between the sensor nodes.

Our approach aims to be as simple and objective as RBF protocol described in [3,4]. However, it is also our goal to provide more accuracy, using packet error rate as a parameter for routing decision. Our solution also differs from [5-7] by using Fuzzy Logic to evaluate the link quality and taking the standard deviation of the measured RSSI to determine the link stability — since thresholds are not defined, nodes with poor link quality may still be used, but only in extreme cases in order to avoid the existence of holes in the network. For [8-11], which are solutions using Fuzzy Logic, we propose a simplification and the use of different parameters, in an approach specially designed for indoor applications.

## 3. FUZZY LOGIC

Traditional logic, based on assumptions and conclusions, cannot represent naturally imprecise phenomena [12], because the propositions come as absolutely true or absolutely false, with no margin for abstract conclusions.

As an alternative to treating problems with inaccurate information or based on incomplete data, we can mention the techniques of Soft Computing. According to [13], Soft Computing seeks to address the uncertainties and inaccuracies inherent in the world we live in order to facilitate the acquisition of more accurate results. Among the techniques of Soft Computing, we can mention Fuzzy Logic, which aims the abstraction of concrete concepts in order to eliminate uncertainties, and evaluate the proposals according to rules based on experience.

In [12] it is stated that the Fuzzy Logic, unlike traditional logic, allows different levels of certainty. This means that it is possible to characterize variables across membership functions that overlap and are influenced by modifiers. Thus, it is possible to deal with abstract concepts similar to human thinking, not considering absolute values, but rather the gradual perception of a particular information.

Therefore, taking into account the uncertainties of the factors influencing the quality of communication in a dynamic environment, we can apply the Fuzzy Logic to verify a link quality index based on the average RSSI, Standard Deviation of the RSSI, and the Packet Loss in order to determine the best routes for packet forwarding.

## 4. PROPOSAL

The proposed algorithm uses Fuzzy Logic to evaluate quality parameters from each link in the WSN. It is analysed the average RSSI level, the standard deviation of the measured RSSI, and the packet error rate in order to classify the link for the routing decision. Once the cost of each link is determined, it uses Dijkstra's algorithm [14] to perform the search of best routes and thereby reconfigure the network.

The base station (sink) acts as a coordinating entity, collecting the RSSI measures, standard deviation, and Packet Error Rate (PER) from each sensor node. Each sensor is responsible for storing the requested data in its own memory and carry out of the necessary measures. However, measurements occur during normal network traffic, not hampering the network overall performance with excessive control packets. The base station is also endowed with a better computational power and radio coverage.

The operating steps of the FLBRA are described as follows:

*Setup Phase* - the base station sends a beacon message to all sensors in the network. The sensors use this message to evaluate their RSSI with the base station. Each sensor node near the base station, and with better RSSI, answers the message. In order to avoid collisions, sensor nodes wait for a time interval based on their RSSI with the base station before sending the answer. Neighbour nodes listen to the answers and keep the RSSI values from other sensors in a local database. The base station waits for the answer of each node in the network until a timeout is expired. After that, the base station sends a new broadcast message asking for RSSI information from the neighbours of each sensor. The sensors answer the base station with the information of their neighbours. At this moment, the base station acknowledges the existence of other sensors, for which the replies could not reach the sink in a single hop. By using the received data, the base station runs the fuzzy logic procedure and the Dijkstra algorithm in order to create the routing table. The base station sends the routing information for each sensor node, which starts forwarding packets according to the new rules. A new broadcast message asking for RSSI information from neighbours is sent in order to identify new sensors. This process is repeated until the base station detects no new sensors in the network.

The Algorithm 1 represents the overall procedures of the setup phase.

**Algorithm 1** NETWORK SETUP

1:    **Function** setup_phase()
2:    send_broadcast(*beacon_message*)   //*sends the beacon message to the sensor nodes*
3:    *timeout* = set_timeout(*null*)   //*starts a timeout waiting for sensor nodes replies*
4:    [*net_info, count_sensors, neighbour_sensors*] = wait_response(*timeout*) //*receives information from sensor nodes*
5:    send_broadcast(*rssi_request, count_sensors*)   //*sends the request for rssi readings*
6:    *timeout* = set_timeout(*count_sensors*)   //*starts a timeout waiting for sensor nodes replies based on the number of sensor nodes*
7:    [*net_info, count_sensors, neighbour_sensors*] = wait_response(*timeout*) //*receives information from sensor nodes*
8:    route_definition(*net_info, count_sensors*)   //*defines the routes based on the received information*
9:    **while** *neighbour_sensors<>null* **do**
10:      send_broadcast(*rssi_request, neighbour_sensors*)   //*sends the request for rssi readings from neighbours*
11:      *timeout* = set_timeout(*neighbour_sensors*)   //*starts a timeout waiting for sensor nodes replies based on the number of sensor nodes*
12:      [*net_info, count_sensors, neighbour_sensors*] = wait_response(*timeout*) //*receives information from sensor nodes*
13:      route_definition(*net_info, count_sensors*)   //*defines the routes based on the received information*
14:    **end**
15:    **return** (*null*)
16:
17:    **Function** wait_response(*timeout*)
18:    **while NOT** expired(*timeout*) **do**   //*keep listening sensor while timout is not expired*
19:      *sensor_info* = receive_answer()   //*receive the answers from each sensor node*
20:      *timeout* = update_timeout()   //*update timeout for each new detected sensor*
21:      [*net_info, count_sensors, neighbour_sensors*] = new_sensor_check(*sensor_info*)   //*number of sensor nodes that can reach the base*
22:    **end**
23:    **return** (*net_info, count_sensors, neighbour_sensors*)
24:
25:    **Function** route_definition(*net_info, count_sensors*)
26:    *path_info* = fuzzy_system(*net_info*)   //*runs the fuzzy system*
27:    *route_table* = dijkstra(*path_info*)   //*runs the Dijkstra algorithm*
28:    **For each** *sensor* **in** *count_sensors*   //*changes route to each sensor node in the network*
29:      change_route(*sensor, route_table*)   //*send the message to change routes*
30:    **end**
31:    **return**(*null*)

*Operation Phase* – once the initial routing table is defined, the network starts the operation phase. The sensors forward their data to the sink following the rules defined in the routing table. At the same time, RSSI information from neighbours and PER are collected and stored for a future check. After a predefined time, the base station sends a new broadcast message asking for new readings. The sensors forward the stored information from their neighbours and the base station analyse if a change in the routing table is required.

The Algorithm 2 represents the operation phase of the FLBRA protocol.

### Algorithm 2 NETWORK OPERATION

```
1:   Function operation_phase()
2:      network_status = OPERATIONAL
3:      timeout = THRESHOLD
4:      while network_status = OPERATIONAL do
5:          [net_info, data] = wait_sensor_data(timeout)  //waits for data from sensor nodes
            until timeout expires
6:          if data <> null then   //forwards data to application layer if any
7:              application_layer(data)
8:          end
9:          if expired(timeout) then  //checks if timeout expired for network status check
10:             network_status = network_check()
11:             if network_status = FAULTY then //checks if network is working fine
12:                 setup_phase()       //in case network needs reconfiguration
13:             end
14:             timeout = THRESHOLD //updates the timeout threshold
15:         end
16:     end
17:     return (null)
18:
19:  Function network_check()
20:     send_broadcast(rssi_request, count_sensors)    //sends the request for rssi readings
21:     timeout = set_timeout(count_sensors)  //starts a timeout waiting for sensor nodes
            replies based on the number of sensor nodes
22:     [net_info, count_sensors, neighbour_sensors] = wait_response(timeout) //receives
            information from sensor nodes
23:     if neighbour_sensors > 0 then   //check if news sensor nodes were detected/added
24:         return(FAULTY)
25:     end
26:     new_path_info = fuzzy_system(net_info)
27:     if new_path_info <> path_info then
28:         return(FAULTY)
29:     end
30:     return(OPERATIONAL)
```

The process to evaluate the cost of each link makes use of fuzzy logic. Hence, it is necessary to define the membership functions of each variable to be analysed (RSSI, Standard Deviation and PER). One must also specify an output variable named "Cost", which is the abstraction of the intervals that are used to determine the cost of the link, where the pertinence of the resulting assessment of the rules will be identified. The sets that comprise the variable "Cost" are: High, Medium and Low as represented in Figure 1. RSSI values were also divided in a membership function and categorized into: Weak, Average and Strong as described in Figure 2. Likewise, the

International Journal of Computer Science & Information Technology (IJCSIT) Vol 6, No 5, October 2014

standard deviation sets were also defined: Good, Average and Bad also represented in Figure 3; and PER variable were represented with Low, Medium and High sets as described in Figure 4. Importantly, the values for the ranges of the variables RSSI and Standard Deviation may vary according to the minimum acceptable power of the radio being used. For this work, we defined a sensitivity of -90 dBm for the purpose of simulation, since it is a practical limit.

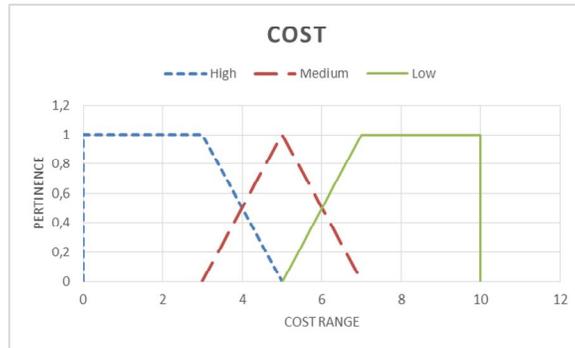

Figure 1 Cost Membership Function

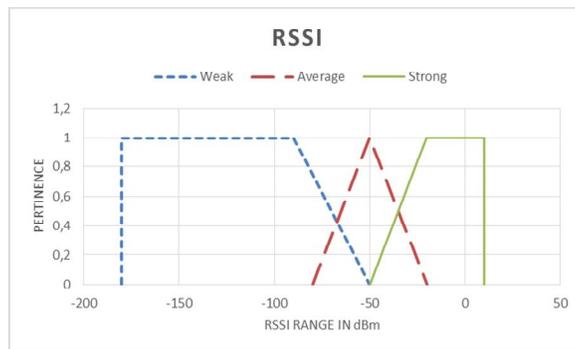

Figure 2. RSSI Membership Function

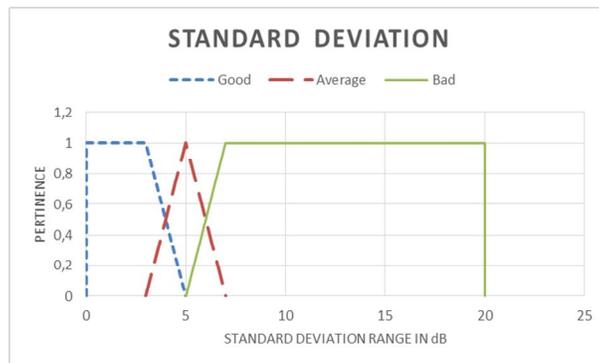

Figure 3.Standard Deviation Membership Function

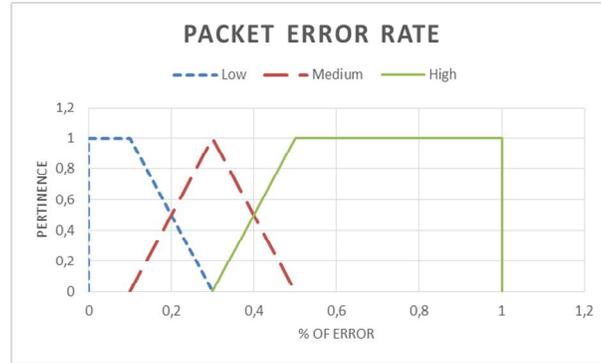

Figure 4.Packet Error Rate Membership Function

Once defined the membership functions of the variables to be analysed, it is necessary to identify the relevance of each value obtained from the measurements in relation to the fuzzy sets. The pertinence defined for each measure of each fuzzy set is used as input for the execution of the defined rules. Table 1 represents the fuzzy rules defined for this work. The execution of the rules allows the identification of subsets in the membership function of the variable "Cost"; in other words, it is possible to verify the areas of influence of each rule for each parameter in the link cost definition. The definition of fuzzy rules is an important step of the process. It reflects the empirical analysis of the studied system. It is possible to check in [15] the Mamdani inference method that was used in this work, which is basically the execution of a IF…THEN process.

Table 1.Fuzzy Rules

| Parameter 1 | Parameter 2 | Parameter 3 | Consequence |
| --- | --- | --- | --- |
| High PER | High PER | High PER | High Cost |
| Medium PER | Medium PER | Medium PER | High Cost |
| Low PER | Weak RSSI | Bad Standard Deviation | High Cost |
| Low PER | Weak RSSI | Average Standard Deviation | Medium Cost |
| Low PER | Weak RSSI | Good Standard Deviation | Low Cost |
| Low PER | Average RSSI | Bad Standard Deviation | High Cost |
| Low PER | Average RSSI | Average Standard Deviation | Medium Cost |
| Low PER | Average RSSI | Good Standard Deviation | Low Cost |
| Low PER | Strong RSSI | Bad Standard Deviation | High Cost |
| Low PER | Strong RSSI | Average Standard Deviation | Low Cost |
| Low PER | Strong RSSI | Good Standard Deviation | Low Cost |

To determine an absolute value (called crisp) for the effective cost of the link it is necessary to implement a defuzzification strategy. The centroid method (centre of area) is used, since it presents lower mean square errors in comparison to other methods, such as mean of maximum method [16]. Once identified the crisp value of the link cost to each sensor node, Dijkstra's algorithm is run in order to find the best routes for packet forwarding. At this point, the routing table of each sensor is defined, and it is identified the sequence to be followed for sending packets from the base station to a specific sensor node and vice versa.

## 5. RESULTS

To evaluate the performance of the proposed algorithm in comparison to the RBF protocol, it was developed a simulator using SciLab [17]. RSSI values were generated taking into account the distance between the sensor nodes in each scenario. For this process, we adapted the model described in [18]. The standard deviation was calculated based on the generated RSSI, and the PER was randomly attributed. The network was designed identically for both FLBRA and RBF solutions.

Six scenarios were defined only varying the total amount of sensor nodes, and the total size of the environment. The distance between the sensor nodes was kept as a constant and it was set at 3 m. Table 2 shows the characteristics of each scenario.

Table 2. Test Scenarios

| Scenario | Quantity of Sensor Nodes | Room size |
|---|---|---|
| S01 | 8 | 36 m² |
| S02 | 24 | 144 m² |
| S03 | 48 | 324 m² |
| S04 | 80 | 576 m² |
| S05 | 120 | 900 m² |
| S06 | 160 | 1296 m² |

To establish a comparison factor between the FLBRA and the RBF protocol, we defined the process presented in (1):

$$F = \sum_{i=1}^{n}(S_{FLBRA\,i} - S_{RBF\,i})/n \, , \, [-1 \leq F \leq 1] \quad (1)$$

Where:

F: single comparison parameter
N: number of sensor nodes
SFLBRA: Success rate based on the proposed fuzzy solution
SRBF: Success rate based on the RBF solution
Success rates can be found as follows in (2.a) and (2.b):

$$S_{FLBRA} = 1 - PER_{FLBRA} \, , \, [0 \leq PER_{FLBRA} \leq 1] \quad (2.a)$$

$$S_{RBF} = 1 - PER_{RBF} \, , \, [0 \leq PER_{RBF} \leq 1] \quad (2.b)$$

Where:

PERFLBRA: packet error rate of a specific link for the solution based on fuzzy logic

PERRBF: packet error rate of a specific link for the solution based on RBF

The parameter F represents a factor of comparison between FLBRA and RBF. F values between [-1,0[ demonstrate that the RBF solution is more efficient. For F = 0, we have that the solutions are similar and break even at the same results. As for F values between ]0,1], we have that FLBRA has a better performance compared to RBF.

For each defined scenario, it was executed 100 iterations of the simulation in order to generate values of the F parameter and evaluate the consistency of the proposed routing solution. Figure 5 shows a graph of the performance based on the evaluation of the F parameter. The θ1 and θ2 curves are respectively the lower and higher values of the confidence interval based on a confidence level of 95% and FM is the mean of the of the F parameter after 100 iterations.

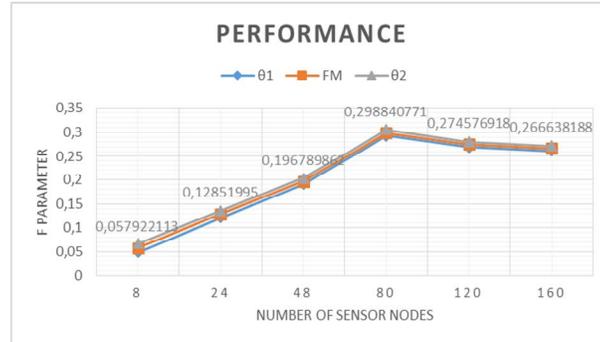

Figure 5. Performance curve (values displayed only for the FM parameter)

It is possible to see that the proposed algorithm always performs better than the strategy based solely on the RSSI. This means that the links created by evaluating the RSSI, standard deviation and packet error rate imply in greater network reliability. It is also observed that as they increased in the number of sensor nodes in the network, we obtain even better results for FLBRA. However, when the number of sensor nodes rises above 100 devices, there is a drop in the performance for both solutions. This is due to the need for more sensor nodes to go on multi hop routes until they can reach the base station. Since there are more sensor nodes and the total area of the test scenario was increased, it is natural that the accumulated performance decreases. The increase of hops works negatively in the packet error probability, as described in (3):

$$\text{PEP} = [1 - \prod_{i=1}^{n}(1 - \text{PER}_i)] \qquad (3)$$

Where:

PEP: final packet error probability

N: number of links

PER: packet error rate of a specific link

FLBRA also performs better in terms of total number of hops for the farthest sensor node. As evidenced by Figure 6, FLBRA uses less hops to reach the base station. This is also observed in Figure 7 which represents the average number of hops in each scenario for both FLBRA and RBF solutions.

International Journal of Computer Science & Information Technology (IJCSIT) Vol 6, No 5, October 2014

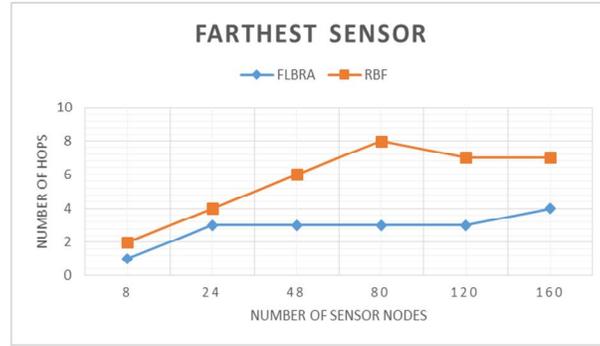

Figure 6 Number of hops for the farthest sensor node

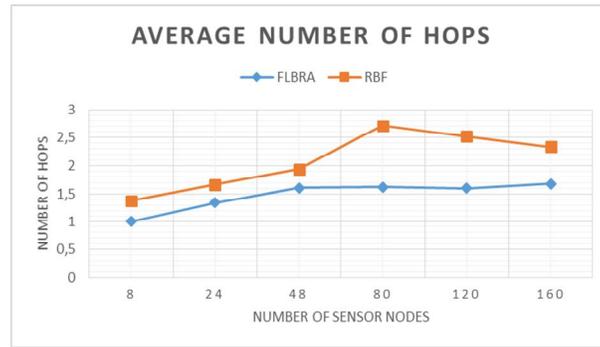

Figure 7.Average Number of Hops

Thus, one can say that FLBRA that is based on the evaluation of RSSI, standard deviation, and packet error rate parameters using fuzzy logic is presented as more appropriate solution for defining routes with better performance, while keeping the simplicity of the implementation.

## 6. CONCLUSION

This paper proposed the use of fuzzy logic as a tool to determine the cost of each link based on signal and communication quality attributes, seeking to identify the best routes in wireless sensor networks, aiming at improving the overall performance of the network in relation to packet error rate.

It can be observed from the simulation that the proposed algorithm presents an improved success rate of packet delivery in relation to the RSSI-based solution. It is also visible that as the amount of sensor nodes in the network raises, the proposal based on fuzzy logic becomes even more advantageous, with larger differences from the RSSI-based routing.

As future work, we intend to modify the algorithm so that it can dynamically vary the intervals of the RSSI, Standard Deviation and PER membership functions in order to adapt to different environments at runtime, while following specified quality criteria. We also intend to develop the algorithm on a real wireless sensor network, using Radiuino [19] platform, and conduct tests in buildings with constant physical variations.

International Journal of Computer Science & Information Technology (IJCSIT) Vol 6, No 5, October 2014## REFERENCES

[1]   D. De Guglielmo and G. Anastasi, "Wireless sensor and actuator networks for energy efficiency in buildings," in Sustainable Internet and ICT for Sustainability (SustainIT), 2012, 2012, pp. 1-3.

[2]   M. S. Aslam, A. Guinard, A. McGibney, S. Rea and D. Pesch, "Wi-design, wi-manage, why bother?" in Integrated Network Management (IM), 2011 IFIP/IEEE International Symposium on, 2011, pp. 730-744.

[3]   A. Awang, X. Lagrange and D. Ros, "RSSI-based forwarding for multihop wireless sensor networks," in The Internet of the FutureAnonymous Springer, 2009, pp. 138-147.

[4]   A. Awang, X. Lagrange and D. Ros, "A cross-layer medium access control and routing protocol for wireless sensor networks," Proc.10emes Journées Doctorales En Informatique Et Réseaux (JDIR 2009), 2009.

[5]   A. Boukerche, H. A. B. F. Oliveira, E. F. Nakamura and A. A. F. Loureiro, "A novel location-free greedy forward algorithm for wireless sensor networks," in Communications, 2008. ICC '08. IEEE International Conference on, 2008, pp. 2096-2101.

[6]   A. Kanzaki, T. Hara, S. Nishio and Y. Nose, "A dynamic route construction method based on measured characteristics of radio propagation in wireless sensor networks," in Advanced Information Networking and Applications (AINA), 2011 IEEE International Conference on, 2011, pp. 30-37.

[7]   Y. Nose, A. Kanzaki, T. Hara, & S. Nishio, "Route Construction Based on Received Signal Strength in Wireless Sensor Networks," (2010) International Conference on Computers and Their Applications (CATA 2010), USA, pp. 127-130.S.

[8]   J. Dastgheib, M. R. S. , H. Oulia, S. J. Mirabedini. A new method for flat routing in wireless sensor networks using fuzzy logic. In Computer Science and Network Technology (ICCSNT), 2011 International Conference on, IEEE Vol. 3, pp. 2112-2116.

[9]   I. Sakthidevi and E. Srievidhyajanani, "Secured fuzzy based routing framework for dynamic wireless sensor networks," in Circuits, Power and Computing Technologies (ICCPCT), 2013 International Conference on, 2013, pp. 1041-1046.

[10]  S. S. Babu, A. Raha, M. K. Naskar, O. Alfandi and D. Hogrefe, "Fuzzy logic election of node for routing in WSNs," in Trust, Security and Privacy in Computing and Communications (TrustCom), 2012 IEEE 11th International Conference on, 2012, pp. 1279-1284.

[11]  N. Baccour, A. Koubâa, H. Youssef, M. B. Jamâa, D. Do Rosário, M. Alves and L. B. Becker, "F-lqe: A fuzzy link quality estimator for wireless sensor networks," in Wireless Sensor NetworksAnonymous Springer, 2010, pp. 240-255.

[12]  L. A. Zadeh, "A summary and update of "Fuzzy logic"," in Granular Computing (GrC), 2010 IEEE International Conference on, 2010, pp. 42-44.

[13]  L. A. Zadeh, "Soft computing and fuzzy logic," Software, IEEE, vol. 11, pp. 48-56, 1994.

[14]  A. S. Tanenbaum, "Computer Networks 4th Edition◦." 2003.

[15]  E. H. Mamdani, "Application of fuzzy algorithms for control of simple dynamic plant," Electrical Engineers, Proceedings of the Institution of, vol. 121, pp. 1585-1588, 1974.

[16]  C. Lee, "Fuzzy logic in control systems: fuzzy logic controller. II," Systems, Man and Cybernetics, IEEE Transactions on, vol. 20, pp. 419-435, 1990.

[17]  S. Enterprises, "Scilab: Free and Open Source software for numerical computation," Scilab Enterprises, Orsay, France, 2012.

[18]  C. Papamanthou, F. P. Preparata, R. Tamassia. "Algorithms for location estimation based on rssi sampling." Algorithmic Aspects of Wireless Sensor Networks. Springer Berlin Heidelberg, 2008. 72-86.

[19]  O. Branquinho, "Plataforma Radiuino para Estudos em Redes de Sensores Sem Fio," http://www.radiuino.cc, Em, pp. 30-09, 2011.
96